\documentclass[twocolumn,english,aps,graphicx,amsmath,showpacs]{revtex4}
\usepackage[T1]{fontenc}
\usepackage[latin1]{inputenc}
\usepackage{babel}
\usepackage{graphics}

\makeatletter

\providecommand{\LyX}{L\kern-.1667em\lower.25em\hbox{Y}\kern-.125emX\@}

\makeatother
\begin{document}

\title{Public-channel cryptography based on mutual chaos pass filters}

\author{Einat Klein\( ^{1} \), Noam Gross\( ^{1} \), Evi Kopelowitz\(
^{1} \), Michael Rosenbluh\( ^{1} \), Lev Khaykovich\( ^{1} \),
Wolfgang Kinzel\( ^{2} \),  Ido Kanter\(^{1} \)}

\affiliation{\( ^{1} \)Department of Physics, Bar-Ilan University,
Ramat-Gan, 52900 Israel,}

\affiliation{\( ^{2} \)Institut f\"ur Theoretische Physik,
Universit\"at W\"urzburg, Am Hubland 97074 W\"urzburg, Germany}

\date{Submitted 28 February 2006}

\begin{abstract}
We study the mutual coupling of chaotic lasers and observe both
experimentally and in numeric simulations, that there exists a
regime of parameters for which two mutually coupled chaotic lasers
establish isochronal synchronization, while a third laser coupled
unidirectionally to one of the pair, does not synchronize. We then
propose a cryptographic scheme, based on the advantage of
mutual-coupling over unidirectional coupling, where all the
parameters of the system are public knowledge. We numerically
demonstrate that in such a scheme the two communicating lasers can
add a message signal (compressed binary message) to the
transmitted coupling signal, and recover the message in both
directions with high fidelity by using a mutual chaos pass filter
procedure. An attacker however, fails to recover an errorless
message even if he amplifies the coupling signal.

\end{abstract}


\maketitle

Coupled semiconductor lasers have been studied extensively in
recent years, due to their inherent non-linearity and chaotic
dynamics. Chaos synchronization for two unidirectionally coupled
lasers, each of which also has self-feedback from an external
cavity, has been demonstrated experimentally and theoretically in
\cite{elsasser96, shore99,locquet01}. In addition, mutually
coupled semiconductor lasers were also studied extensively
\cite{elsasser01,mutual-coupling,Liu1,Rees}, and revealed
different and interesting dynamics. Chaos synchronization has
attracted even more interest recently, because of its potential
application in optical secret communication systems
\cite{chaos-com,Fischer,CPF_paper}.

In this letter the synchronization properties of mutually coupled
versus unidirectionally coupled lasers are analyzed and compared.
We show that there exists a regime in the coupling parameters
space for which the mutually coupled lasers synchronize very well,
yet a unidirectionally coupled laser does not. Thus in an
application in which synchronization is desirable, mutually
coupled lasers have an advantage.

We use this result to propose a secret communication scheme over a
public-channel based on this advantage of mutual coupling over
unidirectional coupling. We emphasize the public channel nature of
the proposed scheme and the advantages this brings to
cryptographic communication systems that make use of chaos
synchronization. For unidirectionally coupled laser, chaos
synchronization cryptography has been based on a private-key
procedure, where the two communicating parties have a common
secret key prior to the communication process, which they use to
encrypt the message they wish to transmit. The unidirectionally
coupled lasers are usually synchronized in a master-slave
configuration, and the secret key generally consists of the
system's parameters \cite{chaos-com}. The system parameters
provide a private key because the two communicating lasers must
have identical (or nearly identical) parameters, or else
synchronization is impossible. In this letter we propose an
all-optical $public$-channel cryptographic system, in which there
is no need to conceal any of the system's parameters, or to
exchange private information prior to the public-channel
communication process.

In our proposed system, the two communicating lasers are mutually
coupled in such a way that they exhibit isochronal
synchronization, in which there is no delay in their synchronized
signals. Stable isochronal synchronization is achieved due to the
self-feedback of each laser as described in \cite{IsoPaper}.
Message encryption is accomplished by adding a low amplitude
binary message to the chaotic laser fluctuations so that the
ensuing transmitted signal still appears to be chaotic and random.
Both lasers are allowed to send simultaneously, independent
messages to each other and the messages are independently
recovered at both ends of the communication line, ensuring
bidirectional information flow. At the receiving end, both lasers
use a chaos pass filter procedure to decode the message from the
received chaotic signal which we will call a "mutual chaos pass
filter" procedure, denoted MCPF.

Communication security is based on the fact that a third laser,
labeled the "attacker", who tries to synchronize himself to the
transmitted signals, is disadvantaged compared to the mutually
coupled lasers, and although he can manage to partly recover the
message, he has considerably more error bits in his recovered
message, and so his eavesdropping attack can be considered
unsuccessful. The use of MCPF provides two novel and distinct
advantages: it is public-key, i.e. it does not require a
secret-key prior to communication, and it allows for simultaneous
bidirectional communication.

\begin{figure}

{\centering \resizebox*{0.5\textwidth}{0.21\textheight}
{{\includegraphics{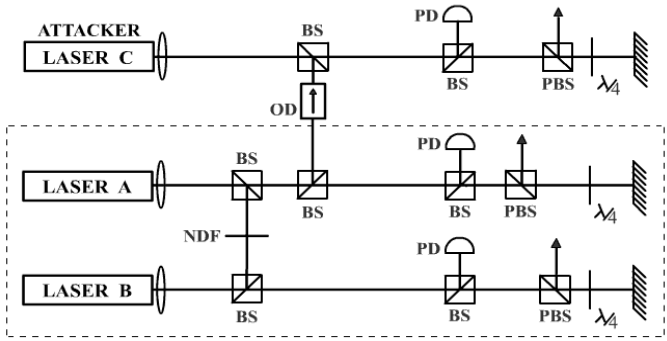}}}
\par}
\caption{\label{schema}Schematic diagram of the experimental
setup. Lasers A and B are mutually coupled, and C, the attacker,
is unidirectionally coupled. BS - beam splitters; PBS -
polarization beam splitters; OD - optical isolator; PD -
photodetectors. }
\end{figure}

\begin{figure}

{\centering \resizebox*{0.4\textwidth}{0.21\textheight}
{{\includegraphics{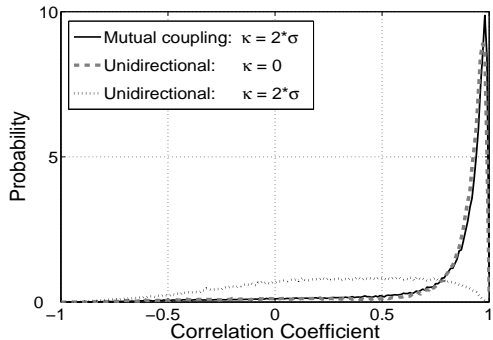}}}
\par}
\caption{\label{hist}Correlation coefficient histograms for
intensity traces divided into 10ns segments, for the cases of
mutual and unidirectional coupling. The total feedback strength is
equal in all cases.}
\end{figure}

Our experimental setup is schematically depicted in Fig.
\ref{schema}, where two mutually coupled external feedback lasers
$A$ and $B$ represent the communicating pair, and the third laser
$C$ which is identically configured but coupled unidirectionally
to one of the pair represents the attacker. In the experiments we
use 3 single-mode lasers, $A$, $B$ and $C$, emitting at 660 nm and
operating close to their threshold. The temperature of each laser
is stabilized to better than 0.01K and all are subjected to a
similar optical feedback. The length of the external cavity is
equal for all lasers and is set to 180 cm (round trip time 12 ns).
The feedback strength of each laser is adjusted using a
$\lambda/$4 wave plate and a polarizing beam splitter. The two
lasers ($A$ and $B$) are mutually coupled by injecting a fraction
of each one's output power to the other. Coupling powers are
adjusted using a neutral density filter. The attacker laser ($C$),
is coupled unidirectionally to one of the mutually coupled lasers,
with unidirectionality ensured by an optical diode (-33 dB). All
the coupling optical paths are set to be equal the self feedback
round trip path. Three fast photodetectors (response time $<$ 500
ps) are used to monitor the laser intensities which are
simultaneously recorded with a digital oscilloscope (500MHz,
1GS/s).

For the case of unidirectionally coupled lasers two types of
synchronization, identical and generalized, have been observed
\cite{TwoSync,openVSclose}. In our experiments we have focused on
the identical synchronization, (which is isochronal) i.e. the two
lasers receive the same total feedback intensities \cite{Remark1}.

We compare mutual versus unidirectional coupling over a range of
coupling and self feedback strengths, denoted $\sigma$ and
$\kappa$ respectively. In the experiment the total feedback
intensity is such that it results in the reduction of the solitary
laser's threshold current by approximately 5$\%$. While keeping
the total feedback of all lasers equal, we vary the values of
$\sigma$ and $\kappa$ over the entire parameter space. The degree
of synchronization between the lasers is evaluated via the
cross-correlation function \cite{IsoPaper}. The time dependent
intensity traces are divided into 10ns segments (containing 10
sample points) and the overlap between matching segments is
calculated and arranged in a histogram. We find that for mutual
coupling robust and stable synchronization is obtained at least
from $\kappa = 2\cdot\sigma$ to $\kappa = 0.5\cdot\sigma$ (see
also ref. \cite{IsoPaper}). For the case of unidirectional
coupling, good synchronization is found for $\kappa = 0$ but it
deteriorates quickly as we increase the part of $\kappa$ in the
total feedback \cite{openVSclose}.

In Fig. \ref{hist} we show a representative point in the parameter
phase space where mutual coupling is advantageous over
unidirectional coupling, by presenting a comparison of overlap
histograms for 3 situations: mutual coupling $\kappa =
2\cdot\sigma$, unidirectional coupling $\kappa = 0$, and
unidirectional coupling $\kappa = 2\cdot\sigma$. While the first
two show good synchronization with a mean value of 0.84, the third
is significantly worse.

\begin{figure}
\vspace{-0.8cm} {\centering
\resizebox*{0.5\textwidth}{0.3\textheight}
{{\includegraphics{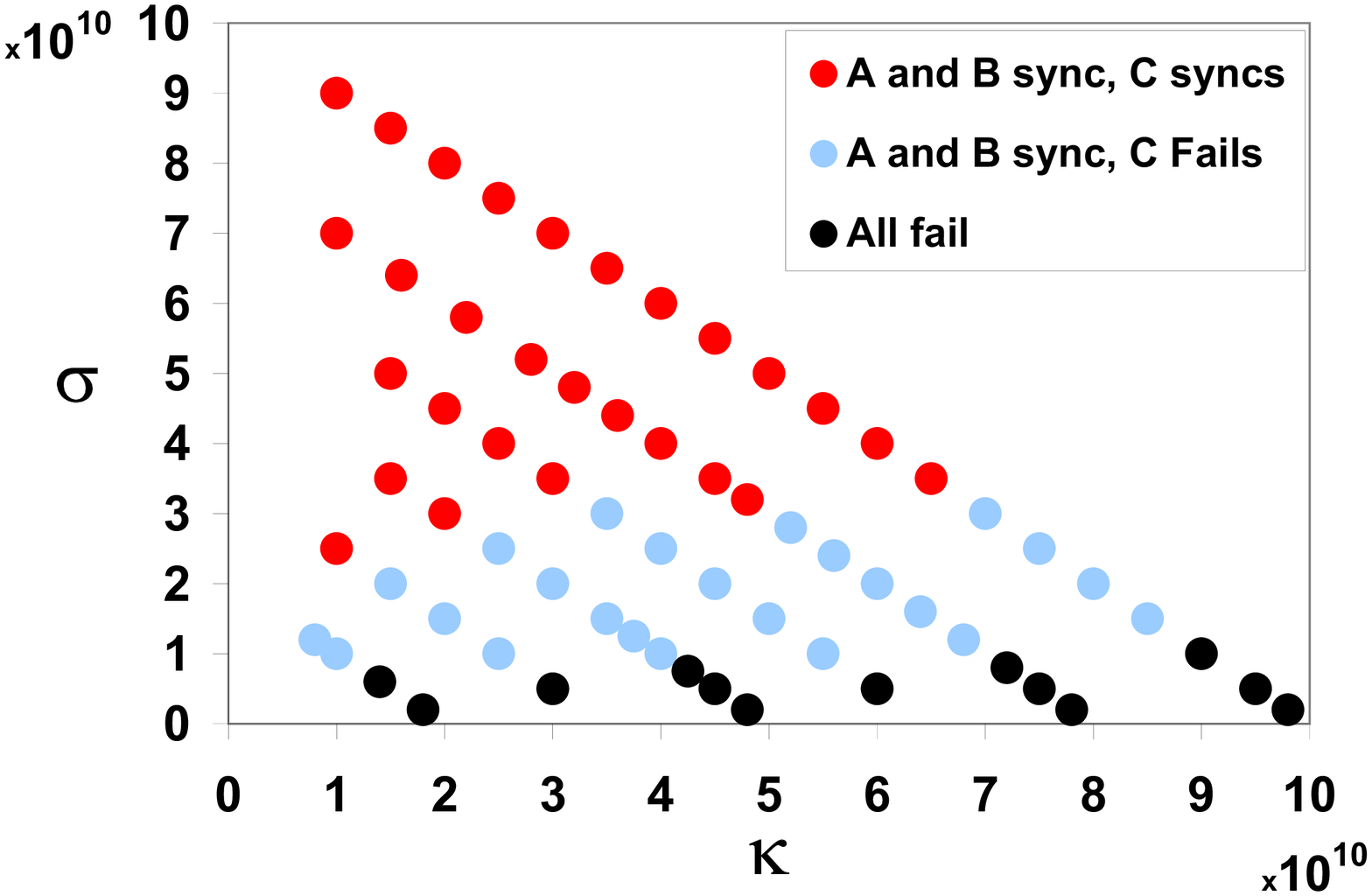}}}
\par}
\vspace{-1.cm} \caption{\label{sim2}Success or failure of
synchronization for the parties and the attacker for a range of
parameter values $\kappa$, feedback strength, and  $\sigma$,
coupling strength.}
\end{figure}

The advantage of mutual coupling over unidirectional is also
obtained in our numerical simulations. We calculate the system
behavior using the Lang Kobayashi differential equations, as
defined in \cite{alhers98} (and also in Eq. [1] below, when taking
$M=0$). For the dynamics parameters we used the values in
\cite{alhers98}, and the two Lasers $A$ and $B$ were found to be
synchronized isochronally \cite{Remark1,IsoPaper}.

We now discuss the two dimensional phase space, defined by
parameters $\kappa$ and $\sigma$ of lasers $A$ and $B$ and the
attacker to either $A$ or $B$. This phase space is characterized
by the following three regimes as depicted in Fig. \ref{sim2}. The
red regime where $\sigma$ is strong enough in comparison to
$\kappa$, and all lasers are synchronized. The black regime where
the coupling is negligible and there is a lack of synchronization
between any of the lasers. Most interesting is the window of the
light blue regime where $A$ and $B$ are synchronized, but $C$
fails to synchronize (we define failure when the average
correlation is less than 0.7). It is in this regime that mutual
coupling is superior to unidirectional coupling, thus even if
laser $C$ uses the exact same parameters as lasers A and B, the
fact that he is not mutually coupled to $A$ or $B$ but only
listening, affects his synchronization ability. This new effect
has been observed in other chaotic systems as well \cite{Losy} and
is at the center of our proposed cryptographic system presented
below. Note that if the attacker amplifies the coupling signal and
uses a stronger $\sigma$ (outside the light blue regime) such as
$\sigma_{attacker} = \kappa+ \sigma$ of of $A$ and $B$, he manages
to synchronize very well. However, in the cryptographic system we
propose, such synchronization does not allow the attacker to
reliably decode the message.

In our proposed MCPF cryptography system, each laser transmits a
signal to the other laser that consists of the original chaotic
laser signal with some added low amplitude message signal. Our
simulations show that synchronization is possible even in the
presence of the added message. In the Lang-Kobayashi equations the
added message is represented by another term $M_{A/B}(t)$ in the
first two equations which is time dependent and different for
lasers $A$ and $B$. The dynamics of laser $A$ are thus given by
the following coupled differential equations for the optical
field, $E$, the optical phase $\Phi$, and the excited state
population, $n$;

\begin{eqnarray}
 & &\frac{dE_{A}}{dt}=\frac{1}{2}G_{N}n_{A}E_{A}(t)+\frac{C_{sp}\gamma[N_{sol}+n
_{A}(t)]}{2E_{A}(t)}\nonumber\\
 & &~~~+ \kappa E_{A}(t-\tau)cos[\omega _{0} \tau + \Phi
_{A}(t)-\Phi_{A} (t-\tau)]\nonumber\\
 & &~~~+ \sigma (E_{B}(t-\tau)+M_B(t-\tau))\nonumber\\
 & &~~~\cdot cos[\omega _{0} \tau +\Phi_{A}(t)-\Phi _{B}(t-\tau)]\\
 & &\frac{d\Phi_{A}}{dt}=\frac{1}{2}\alpha G_{N}n_{A}
\nonumber\\
 & &~~~-\kappa \frac{E_{A}(t-\tau)}{E_{A}(t)}
sin[\omega _{0} \tau +\Phi _{A}(t)-\Phi_{A}
(t-\tau)]\nonumber\\
&  &~~~-\sigma \frac{(E_{B}(t-\tau)+M_B(t-\tau))}{E_{A}(t-\tau)}\nonumber\\
 & &~~~\cdot sin[\omega _{0} \tau +\Phi_{A}(t)-\Phi
_{B}(t-\tau)]\nonumber\\
&  &\frac{dn_{A}}{dt}=(p-1)J_{th}-\gamma n_{A}(t)-[\Gamma +
G_{N}n_{A}(t)]E_{A}^{2}(t)\nonumber
\end{eqnarray}

\noindent and likewise for laser $B$. We obtain that for a wide
range of values of $\kappa$ and $\sigma$ the two lasers achieve
stable isochronal synchronization, despite the fact that each
laser is receiving an additional and different time dependent
message. This message $M_A$, sent from $A$ to $B$, is recovered by
laser $B$ via a chaos pass filter procedure, as $\tilde{M}_A$ by
subtracting his output from the received input in the following
manner: $\tilde{M}_A = E_A + M_A - E_B$, and then averaged over
1ns windows, giving the recovered message as $<\tilde{M}_A>$. The
same method is used by the other laser.

When considering the security of this public cryptographic system
we consider an attacker who is "listening" to the communication
channel and wishes to decipher the secret messages that are
transmitted. When the communicating lasers use coupling strength
values in the "light blue regime" of Fig. \ref{sim2}, an attacker
who uses the same parameters fails to synchronize and hence cannot
recover the message correctly. His best chance to synchronize is
by adjusting his parameters to the red region in Fig. \ref{sim2},
where he amplifies the coupling signal strength $\sigma$ and
weakens his self-feedback strength $\kappa$. In this case he
succeeds in synchronizing, as we already explained above, but he
still fails to decipher the message. The reason for his failure is
that the coupling signal consists both of a fraction of the
laser's signal, $E$, and the message, $M$, and so the attacker
amplifies the message as well as the laser's signal. Hence, the
message/signal ratio for the attacker,
$\frac{M}{(1+\frac{\kappa_{attacker}}{\sigma_{attacker}})<E>}$, is
greater than the message/signal ratio for the parties,
$\frac{M}{(1+\frac{\kappa}{\sigma})<E>}$ \cite{RemarkNS}. This
difference in the message/signal ratio causes the attacker to have
more errors in his recovered message. Therefore when the attacker
amplifies the coupling signal and uses a stronger $\sigma$, he
does not manage to recover the message to the same extent as the
parties and we thus conclude that no matter where in the
$\kappa/\sigma$ phase space the attacker works, he deciphers the
message incorrectly.

\begin{figure}
\vspace{-0.8cm} {\centering
\resizebox*{0.5\textwidth}{0.3\textheight}
{{\includegraphics{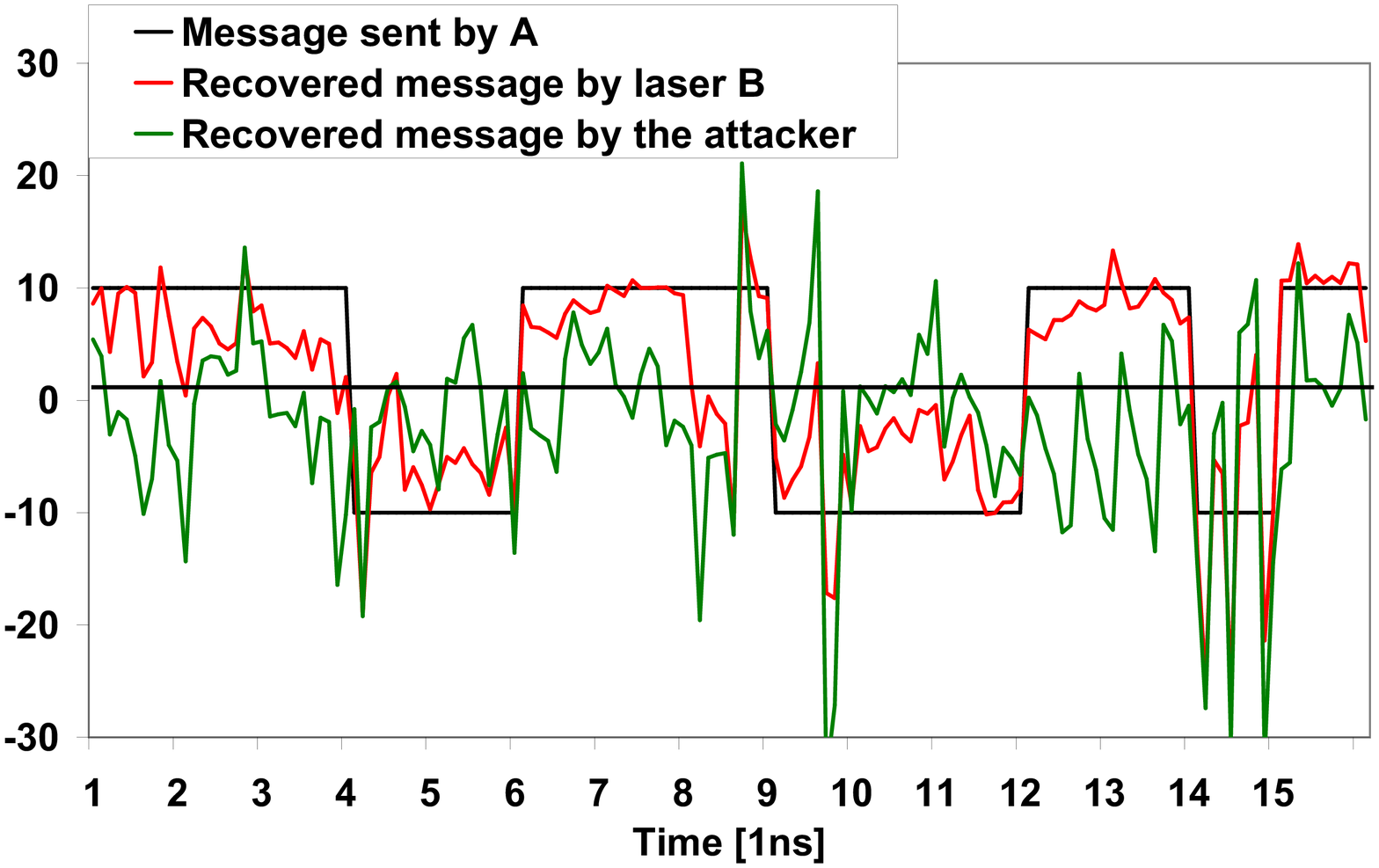}}}
\par}
\vspace{-1.cm} \caption{\label{sim3} A trace of the message
sequence sent from laser A to B (black), that consists of
111001110001101, the recovered message by B (red), and the
recovered message by the attacker (green). The parameters used by
A and B are: $\kappa=11\cdot10^{10}s^{-1}$,
$\sigma=4\cdot10^{10}s^{-1}$. The attacker amplifies the coupling
signal to a maximum and uses $\kappa=0$,
$\sigma=15\cdot10^{10}s^{-1}$.}
\end{figure}

Fig. \ref{sim3} displays the traces of the message
sent by $A$, the message recovered by $B$, and the message recovered
by the attacker, for an attacker using a maximal coupling signal
$\sigma$. One can observe that the attacker's recovered message,
although generally following the original message, has several
mistakes even in this short sequence. When sending compressed data,
even several mistakes can corrupt the entire message. An accepted
measure of the ability to recover a message successfully is the Bit
Error Rate (BER), the probability of a bit to be decoded erroneously
\cite{RemarkBER}.

We measured the BER of the communicating lasers for many different
points in the "light blue regime", for which there is a strong
self-feedback. For every point we examined we looked for the
"best" possible $\kappa,\sigma$ values for the attacker, in the
entire space $(\kappa,\sigma)$, that gives him the minimal BER
value \cite{remark2}. For all the parameters we checked, the BER
of lasers $A$ and $B$ was considerably smaller than the BER of the
best attacker. For example:

\vspace{0.4cm}
\begin{tabular}{|c|c|c|}
 \hline
    & Example 1 & Example 2 \\
 \hline
  $\kappa$ & $8\cdot10^{10} s^{-1}$ & $11\cdot10^{10}s^{-1}$ \\
 \hline
  $\sigma$ & $2\cdot10^{10}s^{-1}$ & $4\cdot10^{10}s^{-1}$ \\
 \hline
  $\frac{M_{A/B}}{<E_{A/B}>}$ & $3\%$ & $12\%$ \\
 \hline
  $\kappa_{attacker}$ & $3\cdot10^{10}s^{-1}$ & $7\cdot10^{10}s^{-1}$ \\
 \hline
  $\sigma_{attacker}$ & $9\cdot10^{10}s^{-1}$ & $8\cdot10^{10}s^{-1}$ \\
 \hline
  $BER_{A/B}$ & $\sim10^{-4}$ & $\sim10^{-4}$ \\
 \hline
  $BER_{attacker}$ & $\sim10^{-2}$ & $\sim10^{-1}$ \\
  \hline
\end{tabular}

\vspace{0.4cm}

\noindent where $<E_{A/B}>$ is the average amplitude of the laser
signal, and the $\kappa_{attacker}$ and $\sigma_{attacker}$
appearing in the table are the values that minimize the BER of the
attacker. The added message was changed randomly every 1ns, giving
a rate of 1Gbits/s. It is clear that the BER of the attacker is
few orders of magnitude higher than the BER of the parties. Hence,
while a compressed block (for instance of 10Kbits) is recovered
properly with high probability by the parties, the attacker has
many ($O(10^3)$) error bits. By reducing the transmission rate,
for instance, and transmitting a bit every 3ns (a 1/3 Gbit/s
rate), the BER of the parties can be reduced, and the gap between
the $BER_{A/B}$/$BER_{attacker}$ is enhanced.

Note that if the two communicating lasers use the same $\kappa$
and $\sigma$ as the optimal attacker in the examples above
($\kappa=3\cdot10^{10}s^{-1}$, $\sigma=9\cdot10^{10}s^{-1}$ or
$\kappa=7\cdot10^{10}s^{-1}$, $\sigma=8\cdot10^{10}s^{-1}$), they
also have BER $\sim10^{-2}$. Hence working in the light blue
regime provides two advantages: the BER in this region is
considerably smaller, and the mutual coupling is superior to the
unidirectional.

We checked that the synchronization of the mutually coupled lasers
is robust under different message signals, such as a "noisy"
message, which was also recovered correctly using the MCPF.
Finally, we considered an attacker who listens to $both$
communication directions, using the coupling signal of both $A$ to
$B$ and $B$ to $A$, and found this attacker to be unsuccessful.

To conclude, we have presented a public-channel cryptographic
system, based on two mutually coupled lasers in a MCPF procedure.
Encryption security is based on the experimentally demonstrated
advantage of mutual coupling over unidirectional coupling. This
system is novel in several aspects: It is an optical communication
system that is public and does not require a secret-key, it
enables two-directional message flow, and the security is not
limited by a small key-space, as in the case of unidirectional
CPF. The system proposed here opens a manifold of possibilities,
for instance: the extension to generate secret communication among
a group of more than two lasers.


\vspace{-0.7cm}
\begin{thebibliography}{9}
\vspace{-1.8cm}
\bibitem{elsasser96} I. Fisher, G.H.M. van Tartwijk,
A.M. Levine, W. Els\"{a}sser, E.O. G\"{o}bel, and D. Lenstra,
Phys. Rev. Lett. {\bf 76}, 220 (1996).

\bibitem{shore99} S. Sivaprakasam and K.A. Shore, Opt. Lett. {\bf 24},
466 (1999).

\bibitem{locquet01} A. Locquet, F. Rogister, M. Sciamanna, P. Megret,
and M. Blondel, Phys. Rev. E {\bf 64}, 045203(R) (2001).


\bibitem{elsasser01} T. Heil ,I. Fisher, W. Els\"{a}sser, J. Mulet and
C. Mirasso, Phys. Rev. Lett. {\bf 86}, 795 (2001);


\bibitem{mutual-coupling} A. Hohl, A. Gavrielides, T. Erneux, and V.
Kovanis, Phys. Rev. A. {\bf 59}, 3941 (1999); J. Mulet C.
Masoller, and C.R. Mirasso \textit{ibid} {\bf 65}, 063815 (2002);
J. Javaloyes, P. Mandel and D. Pieroux, Phys. Rev. E {\bf 67},
036201 (2003); N. Fujiwara, Y. Takiguchi and J. Ohtsubo, Opt.
Lett. {\bf 28}, 1677 (2003).

\bibitem{Liu1} R. Vicente, S. Tang, J. Mulet, C.R. Mirasso and
J.M. Liu, Phys. Rev. E {\bf 70}, 046216 (2004).

\bibitem{Rees}P. Rees, P.S. Spencer, I. Pierce, S. Sivaprakasam, and K.A. Shore,
Phys. Rev. A {\bf 68}, 033818 (2003).

\bibitem{chaos-com} L.M. Pecora and T.L. Carroll, Phys.
Rev. Lett. {\bf 64}, 821 (1990); J.-P. Goedgebuer, L. Larger, and
H. Porte, Phys. Rev. Lett. {\bf 80}, 2249 (1998); G.D. VanWiggeren
and R. Roy, Science {\bf279}, 1198 (1998).

\bibitem{Fischer}A. Argyris, D. Syvridis, L. Larger, V. Annovazzi-Lodi, P. Colet,
I. Fischer, J. Garcia-Ojalvo, C. R. Mirasso, L. Pesquera and K. A.
Shore, Nature 438, 343-346 (2005).


\bibitem{CPF_paper} I. Fischer, Y. Liu and P. Davis, Phys. Rev. A {\bf 62}, 11801 (2000).


\bibitem{IsoPaper}E. Klein, N. Gross, M. Rosenbluh,
W. Kinzel, L. Khaykovich and I. Kanter. (To appear in Phys. Rev.
E.)



\bibitem{alhers98} V. Alhers, U. Parlitz, and W. Lauterborn, Phys. Rev.
E {\bf 58}, 7208 (1998); I.V. Koryukin and P. Mandel, Phys. Rev. E
{\bf 65}, 026201 (2003).


\bibitem{TwoSync}Y. Liu, P. Davis, Y. Takiguchi, t. Aida, S. Saito, and J. Liu, IEEE Journal of Quantum Electronics, vol.
39, no. 2 (2003).

\bibitem{openVSclose}R. Vicente, T. Perez, and C.R. Mirasso, IEEE Journal of Quantum Electronics, vol.
38, no. 9 (2002).

\bibitem{Remark1} For the appearance of the second type (generalized
synchronization), high coupling intensities are required, which is
a disadvantage for the attacker (laser $C$).


\bibitem{Losy}E. Klein, R. Mislovaty, I. Kanter and W. Kinzel, Phys.
Rev. E  {\bf 72}, 016214 (2005).


\bibitem{RemarkNS}If the attacker wishes to keep the same
message/signal ratio,
$\frac{\kappa_{attacker}}{\sigma_{attacker}}=\frac{\kappa}{\sigma}$,
then if $\kappa$ and $\sigma$ are in the light blue regime of Fig.
3, so must be $\kappa_{attacker}$ and $\sigma_{attacker}$.

\bibitem{RemarkBER}Note that minimizing the BER or maximizing the
synchronization are in general two different tasks. Minimizing the
BER for instance, requires that the average recovered signal over
a window, have the same sign as the message, while the local
synchronization can be reduced.

\bibitem{remark2}Since we are working in the LFF regime, we
excluded the downfalls from the BER calculation (for all lasers),
since during the downfalls the lasers temporarily fall out of
synchronization.


\end{thebibliography}
\end{document}